\documentclass[aps,prb,showpacs, amsmath,amssymb,floatfix,groupedaddress,twocolumn,superscriptaddress]{revtex4-2}
\usepackage{graphicx}
\usepackage{dcolumn}
\usepackage{bm}
\usepackage{times}
\usepackage[usenames,dvipsnames]{color}
\usepackage{colordvi,epsf}
\usepackage{multirow}
\usepackage{longtable}
\usepackage[utf8]{inputenc}
\usepackage[abs]{overpic}
\usepackage{colortab}
\usepackage{soul} 

\usepackage{changes}
\definechangesauthor[name=FN, color=blue]{fn}
\newcommand{\uudd}{$uudd$ }
\newcommand{\foursite}{four-site-four-spin }
\newcommand{\mgkm}{$\overline{\rm M}$-$\overline{\Gamma}$-$\overline{\rm K}$-$\overline{\rm M}$}
\newcommand{\gm}{$\overline{\Gamma}$-$\overline{\rm M}$ }
\newcommand{\gk}{$\overline{\Gamma}$-$\overline{\rm K}$ }
\newcommand{\qm}{$\mathbf{q}_{\rm M}$ }
\newcommand{\qk}{$\mathbf{q}_{\rm K}$ }

\newcommand{\fig}{Fig.~}

\usepackage{titlesec}

\titleformat{\section}[display]{\normalfont\bfseries}{}{0.1em}{}[]
\titlespacing{\section}{0.0em}{-0.8em}{0.0em}
\titleformat{\subsection}[display]{\normalfont\bfseries}{}{0.1em}{}[]
\titlespacing{\subsection}{0.0em}{-0.8em}{0.0em}

\usepackage{xcolor}

\begin{document}

\title{
Antiferromagnetic order of topological orbital moments in atomic-scale skyrmion lattices
}
 
\author{Felix Nickel}
\email[Email: ]{nickel@physik.uni-kiel.de}
\affiliation{Institut f\"ur Theoretische Physik und Astrophysik,
Christian-Albrechts-Universit\"at zu Kiel, D-24098 Kiel, Germany}
\author{Andr\'e Kubetzka}
\affiliation{Department of Physics, University of Hamburg, Jungiusstraße 11, 20355 Hamburg, Germany}
\author{Mara Gutzeit}
\affiliation{Institut f\"ur Theoretische Physik und Astrophysik,
Christian-Albrechts-Universit\"at zu Kiel, D-24098 Kiel, Germany}
\author{Roland~Wiesendanger}
\affiliation{Department of Physics, University of Hamburg, Jungiusstraße 11, 20355 Hamburg, Germany}
\author{Kirsten von Bergmann}
\email{kirsten.von.bergmann@physik.uni-hamburg.de}
\affiliation{Department of Physics, University of Hamburg, Jungiusstraße 11, 20355 Hamburg, Germany}
\author{Stefan Heinze}
\email[Email: ]{heinze@physik.uni-kiel.de}
\affiliation{Institut f\"ur Theoretische Physik und Astrophysik,
Christian-Albrechts-Universit\"at zu Kiel, D-24098 Kiel, Germany}
\affiliation{Kiel Nano, Surface, and Interface Science (KiNSIS), University of Kiel, Germany}

\date{\today}

\begin{abstract}
Topological orbital moments can arise in non-coplanar spin
structures even in the absence of spin-orbit coupling and a net topological orbital magnetization occurs for the triple-Q state and for isolated skyrmions.
For atomic-scale skyrmion lattices, 
a significant effect
can also be expected, 
however, no studies have been reported yet.
Here, we observe via spin-polarized scanning tunneling microscopy
a non-coplanar atomic-scale spin structure with a
nearly square magnetic unit cell 
for a pseudomorphic Fe 
monolayer on three atomic Ir layers on the Re(0001) surface. Employing density functional theory (DFT) calculations we consider different skyrmionic lattices 
to find the magnetic ground state. By mapping the 
DFT total energies to an atomistic spin model we demonstrate that these spin textures are 
stabilized by the interplay of the Dzyaloshinskii-Moriya and four-spin interactions. We evaluate the emerging phenomena of the different non-coplanar magnetic states and find significant local topological orbital moments oriented perpendicular to the surface, which order 
in an antiferromagnetic fashion.
\end{abstract}

\maketitle

\section{Introduction} 
Magnetic skyrmions have raised 
widespread attention because of their fascinating topological and dynamical 
properties \cite{Nagaosa2013,Fert2013}. Due to the spin
topology of a skyrmion an emergent magnetic field arises which
causes the topological Hall effect \cite{Taguchi2001}
allowing electrical detection
of single skyrmions~\cite{Maccariello2018}. Another key consequence of electron motion in the fictitious magnetic field is the topological orbital moment. It can occur in 
non-coplanar spin structures even in the absence of spin-orbit coupling
and depends on the local scalar spin chirality \cite{Taguchi2001,Hanke2016,dosSantos2016,Grytsiuk2020}.
The scalar spin chirality is proportional to the topological
charge density and can serve as a measure of
topological transport properties \cite{Taguchi2001,dosSantos2016}.

Topological orbital moments have been predicted
for several types of spin structures such as multi-Q states 
\cite{Hanke2016,Haldar2021}, atomic-scale spin 
lattices \cite{Hoffmann2015}, and isolated
skyrmions \cite{dosSantos2016,Lux2018}. 
In the triple-Q state \cite{Kurz2001,Spethmann2020,Nickel2023}
-- exhibiting tetrahedron angles between adjacent spins on a hexagonal lattice and
a net vanishing spin moment --
the topological orbital moments align such that a net orbital
magnetization remains \cite{Hanke2016}.
Such topological orbital ferromagnetism has also
been proposed for the bulk antiferromagnet 
$\gamma$-FeMn \cite{Hanke2017}.
A net topological orbital moment remains as well for single skyrmions
in a ferromagnetic background
and it has been suggested that these are
observable via XMCD \cite{dosSantos2016}. 
The orbital
magnetization can be manipulated by an external magnetic field which 
has been shown via
the spontaneous topological Hall 
effect in the triple-Q state \cite{Park2023,Takagi2023}.
Other prime candidates for the investigation of topological orbital moments are zero-field non-coplanar magnetic states that have been experimentally found in Fe monolayers in contact with a hexagonal Ir film \cite{Heinze2011,Kubetzka2020}. Due to the large tilting angles between adjacent spin moments such spin textures could result in significant topological orbital moments.

Here, we investigate the non-coplanar spin structure of
Fe/Ir-3/Re(0001) using SP-STM experiments with different tip magnetization directions as well as DFT calculations and an atomistic 
spin model. We show that multi-Q states and atomic-scale skyrmionic lattices
constructed based on the experimentally determined magnetic unit cell
are energetically favored with respect to single-Q
(spin spiral) states. The superposition states are stabilized by
the four-site four spin interaction as shown via mapping DFT total
energies to an atomistic spin model. Due to their non-coplanar spin 
textures 
significant local topological orbital moments arise in these
multi-Q states and atomic-scale skyrmionic lattices.
Intriguingly, these considered spin structures
exhibit antiferromagnetic
order of the arising topological orbital moments. 

\begin{figure}
\includegraphics[width=0.95\linewidth]{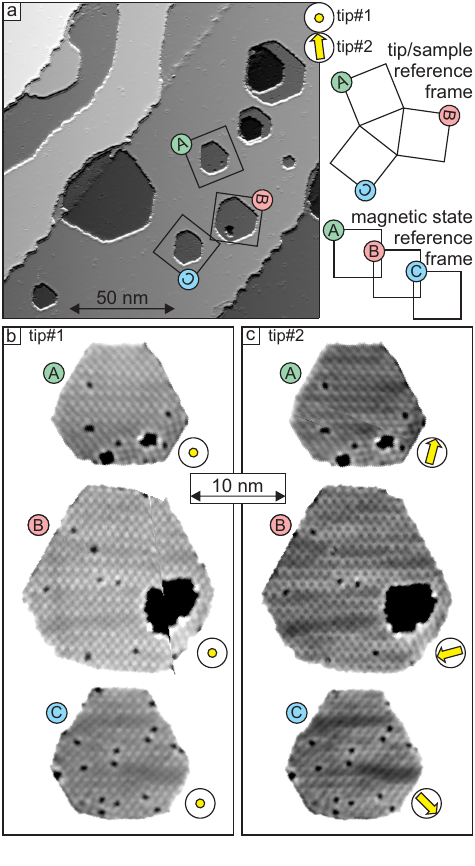}
\caption{\textbf{SP-STM measurements of Fe/Ir-3/Re(0001). }\textbf{a}~Overview partially differentiated constant-current SP-STM measurement. \textbf{b},\textbf{c}~SP-STM measurements of the three Fe/Ir3 areas indicated by squares in \textbf{a}, measured with two slightly different tips as indicated; the gray scale spans 40 pm; the yellow arrows refer to tentative tip magnetization directions and their relative orientation for the rotational magnetic domains. Measurement parameters: \textbf{a}~$U=+100$\,mV, $I=1$\,nA; \textbf{b}~$U=+40$\,mV, $I=1$\,nA; \textbf{c}~$U=+5$\,mV, $I=4$\,nA; all: $T=4$\,K, Cr bulk tip; the glitch in island B of panel \textbf{b} stems from a briefly retracted tip and subsequent creep; the fast scan direction of the islands in \textbf{b} and \textbf{c} is along the horizontal axis of the overview image in \textbf{a}.}
\label{fig:exp}
\end{figure}

\section{Results}
\textbf{SP-STM experiments.}
Pseudomorphic Fe monolayers on Ir ultra-thin films on Re(0001) have been investigated experimentally before~\cite{Kubetzka2020} and in the case of an Ir film with a thickness of three atomic layers a roughly square magnetic unit cell containing about 8 Fe atoms has been found. An overview SP-STM image is shown in Fig.~\ref{fig:exp}a and three Fe monolayer patches are labelled A,B,C. These three Fe islands are embedded in an additional Ir layer and the observed modulations are of magnetic origin. Due to the incompatibility
of the roughly square symmetry of the magnetic state and the hexagonal atomic arrangement three rotational domains of the magnetic state can occur, and the areas labelled by A,B,C correspond to those rotations. In \fig\ref{fig:exp}b and \fig\ref{fig:exp}c the three Fe islands are shown in the magnetic unit cell reference frame, i.e., their relative $120^{\circ}$ rotations on the sample have been accommodated for (cf.\ sketches to the right of \fig\ref{fig:exp}a). Each set of these SP-STM images has been measured with one magnetic tip, but the tip changed between the data displayed in \fig\ref{fig:exp}b and \fig\ref{fig:exp}c. In the images measured with tip\#1 (\fig\ref{fig:exp}b) a square unit cell can be seen clearly in some parts of the islands, however, in other parts of the island a $\textrm{c}(\sqrt{2}\times\sqrt{2})$ superstructure dominates. We attribute this to a variation of the size or the location of the magnetic unit cell with respect to the atomic lattice, either due to an incommensurate magnetic unit cell or due to confinement effects.

Close inspection of the magnetic contrast shows that it is qualitatively similar in all three images of Fig.~\ref{fig:exp}b, whereas the observed magnetic pattern changes for the rotational domains displayed in \fig\ref{fig:exp}c. This is characteristic for a non-collinear magnetic state imaged with an out-of-plane magnetized tip in the first case (\fig\ref{fig:exp}b), and a magnetic tip that also has an in-plane magnetization component in the latter case (\fig\ref{fig:exp}c)~\cite{Heinze2011}. Tentative tip magnetization directions are indicated in yellow to visualize the origin: the sample's out-of-plane components always show the same pattern, regardless of the orientation of the magnetic unit cell; however, an in-plane magnetized tip will pick up different sample magnetization components depending on how the square unit cell is rotated relative to the tip magnetization direction. For tip\#2 a two-dimensionally periodic pattern is observed in islands A and B, whereas for island C a stripe pattern is found, which is reminiscent of the SP-STM measurements of the Fe monolayer on Ir(111), which exhibits a square nanoskyrmion lattice albeit with nearly twice as many atoms in its magnetic unit cell~\cite{Heinze2011}.

\begin{figure}[h]
\includegraphics[width=0.99\linewidth]{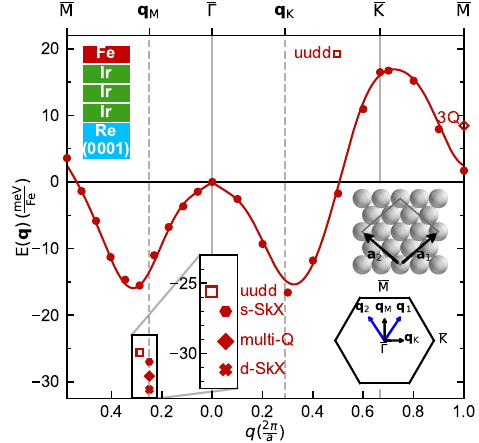}
\caption{\textbf{DFT total energies of spin states in Fe/Ir-3/Re(0001). }
Energy dispersion of spin spirals obtained via DFT
along the high-symmetry lines \mgkm~including SOC
displayed by filled circles. The solid line represents a fit of 
the atomistic model to the DFT values.  
The total energies of the two \uudd states and the 3Q state calculated via DFT are marked. The self-consistently calculated DFT total energies including SOC for the three spin lattices (\fig \ref{fig:unitCell}) are also marked.  In the inset the unit cell of the magnetic contrast is shown with the magnetic unit cell vectors $\mathbf{a}_1$ and $\mathbf{a}_2$ obtained based on the experimental SP-STM images.
The Brillouin zone is also sketched as an inset and the spin spiral vectors $\mathbf{q}_1$, $\mathbf{q}_2$, \qk and \qm are indicated.}
\label{fig:mgkm}
\end{figure}

\textbf{First-principles calculations.}
To understand the origin of the magnetic ground state of Fe/Ir-3/Re(0001), we have performed first-principles calculations based on DFT.
(for computational details see Methods).
We calculate the total energy of various collinear and 
non-collinear magnetic states including spin-orbit coupling (SOC). 
We start with spin spiral states since these represent the general solution 
of the classical Heisenberg model on a periodic lattice
and therefore allow to scan a large
part of the magnetic phase space. The energy dispersion 
$E(\mathbf{q})$ of spin spirals in Fe/Ir-3/Re(0001) (\fig \ref{fig:mgkm}) 
including the effect of SOC, i.e.~Dzyaloshinskii-Moriya interaction (DMI), 
displays two minima 
along the high-symmetry directions \gm and \gk 
with an energy of about 15 meV/Fe atom below the ferromagnetic 
state ($\overline{\Gamma}$ point) 
exhibiting a similar length of 
the spin spiral vector $\mathbf{q}$. 
The corresponding cycloidal spin spirals are stabilized by 
frustrated exchange interactions and DMI (see Supplementary Note 1 and Supplementary
Figure 1 for separate contributions and Supplementary Table 1 for all interaction constants).
Note, that the total
energy scale of the dispersion is below 40 meV/Fe atom
which is an extremely small value. This is due to a small ferromagnetic nearest-neighbor (NN) exchange interaction which
competes with 
antiferromagnetic beyond-NN exchange
(values are given in the Supplementary Table S1).

\begin{figure*}
\includegraphics[width=0.95\linewidth]{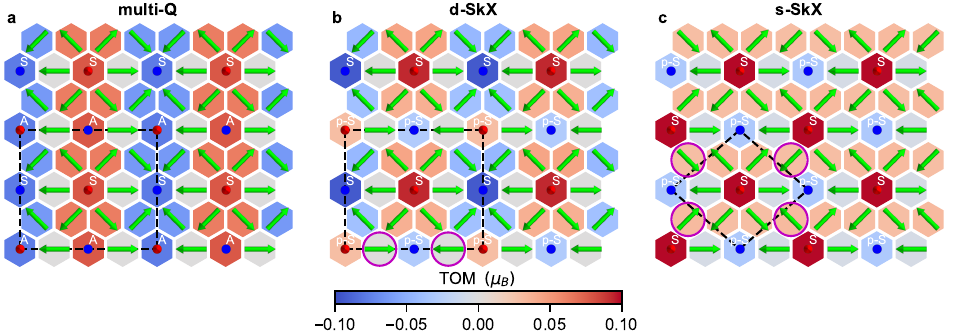}
\caption{\textbf{Spin structure and topological orbital moments
of atomic-scale spin lattices.} \textbf{a} multi-Q state , \textbf{b} double-skyrmionic lattice (d-SkX), and \textbf{c} single-skyrmionic lattice (s-SkX). The multi-Q state 
is constructed from a superposition of spin spirals with $\mathbf{q}$ vectors corresponding to the magnetic unit cell obtained in experiment
(see text for details). The d-SkX lattice  has two spins in the bottom row flipped (see purple circles in \textbf{b})
to optimize the DMI energy.  For the s-SkX lattice  four further spins have been flipped around the x-axis
(see purple circles in \textbf{c}) with 
respect to the multi-Q state  to reduce the size of the unit cell.
The dashed black boxes mark the magnetic unit cell of each lattice. The color of the hexagons display the topological orbital moment (TOM) for each atom calculated by DFT (see color bar at bottom). Red (blue)
color denotes upwards (downwards) pointing TOM with respect to the surface normal.
For the ease of discussion in the text, the centers of atomic-scale substructures 
of the spin lattices are denoted by letters: skyrmion (S), antiskyrmion (A), and pseudo-skyrmion (p-S).}
\label{fig:unitCell}
\end{figure*}

The influence of higher-order exchange interactions (HOI) can be revealed by calculating the total energy of superpositions of spin spirals,
so-called multi-Q states: the $uudd$ states along both high-symmetry directions \cite{Hardrat2009,Kronelein2018,Hoffmann2020,Gutzeit2021} and the 3Q state \cite{Kurz2001,Spethmann2020,Haldar2021,Nickel2023}. Within the Heisenberg model of pair-wise exchange these states are degenerate with the spin spiral (single-Q) states from which they are constructed.  This degeneracy can be lifted by HOI. Therefore, a total DFT energy difference between single-Q and multi-Q states 
is an indication of 
HOI. 
In \fig \ref{fig:mgkm} a clear deviation of the energies of the \uudd and the 3Q states from the corresponding 1Q states is evident, which means that HOI exhibit a significant strength in this system.
Out of the spin spiral states and the mentioned model-type multi-Q states the \uudd state in $\overline{\mathrm{\Gamma M}}$ direction has the lowest energy, with a $\Delta E$ of $10$\,meV lower than the minima of the single-Q state dispersion.

To construct a superposition state with the experimentally observed magnetic unit cell (cf.~\fig \ref{fig:exp}), we convert the real space magnetic lattice vectors of the roughly square 
commensurate unit cell $\mathbf{a}_1$ and $\mathbf{a}_2$ into spin spiral vectors $\mathbf{q}_1$ and $\mathbf{q}_2$ in reciprocal space (see inset of Fig.~\ref{fig:mgkm}). From $\mathbf{q}_1$ and $\mathbf{q}_2$ we can obtain spin spiral vectors \qm and \qk in the high symmetry directions.
These are given by $\mathbf{q}_{\rm M} = \frac{1}{2}(\mathbf{q}_1+\mathbf{q}_2)=
\frac{\pi}{2a}\hat{e}_y$ and $\mathbf{q}_{\rm K} = \frac{1}{2}(\mathbf{q}_1-\mathbf{q}_2)=\frac{\pi}{\sqrt{3}a}\hat{e}_x$, respectively. All these $\mathbf{q}$ vectors are close to the energy minima of $E(\mathbf{q})$ (Fig.~\ref{fig:mgkm}).

The multi-Q state is constructed from these spin spirals using the analytical expression given in Ref.~\cite{Heinze2011} and the spin texture is displayed in Fig.~\ref{fig:unitCell}a (For the construction of the multi-Q state see Supplementary Note 2).
Within DFT it is energetically significantly
lower in total energy than the spin spiral minimum and 
a few meV/Fe atom lower than the $uudd$ state along
$\overline{\Gamma {\rm M}}$ (Fig.~\ref{fig:mgkm}). 
The two-dimensional magnetic unit cell of the multi-Q state
contains 16 atoms (Fig.~\ref{fig:unitCell}a) and two spins point upwards with respect to the surface (red)
and two spins point downwards (blue). 
All other spins are oriented in the film plane (green arrows).  Each of the upward or downward pointing spins can be viewed as the center of an atomic-scale skyrmion or antiskyrmion building block of the lattice, marked in Fig.~\ref{fig:unitCell}a with S and A, respectively. We denote as a skyrmion substructure
a plaquette consisting of a central out-of-plane spin and six neighboring spins pointing either towards or away from the central spin. The antiskyrmion 
plaquette consists of a central out-of-plane spin and six neighbors
with two spins pointing outwards or inwards and the remaining four spins pointing nearly perpendicular to the connecting vector to the central spin. The profile of the in-plane component of these spins coincides with the profile of an antiskyrmion \footnote{When inspecting the six nearest neighbors, the DMI contribution is not vanishing as in an isolated antiskyrmion. The name of the substructure reflects here the magnetisation direction.}. 

For each of these substructures two versions exist, one with the central spin pointing downward, referred to as down-S or down-A, and one with the central spin pointing up, referred to as up-S or up-A. Because of their atomic-scale size, neighboring building blocks share several spins. In the multi-Q state horizontal rows of up-S and down-S are separated by rows of down-A and up-A.

\textbf{Topological orbital moments.}
In order to gain further insight into the properties of the multi-Q state we have calculated the topological orbital moments (TOM)
via DFT. In \fig \ref{fig:unitCell}a the TOM of every Fe atom is displayed by the color of the hexagon at the corresponding lattice site (for values see Fig.~S3). Note, that up-S possess a positive TOM, while down-S exhibit a negative TOM and vice versa for up- and down-A. Therefore, in the multi-Q state the sites with upward (red) and downward (blue) pointing TOM form rows perpendicular to the rows of skyrmions and antiskyrmions. 
The total TOM per skyrmion or antiskyrmion plaquette amounts to about 
0.32~$\mu_{\rm B}$, while the sum of all TOM in the magnetic unit cell vanishes.

Next to the multi-Q state also other spin configurations within the experimentally found magnetic unit cell are possible and we will consider two other likely non-coplanar candidates.
By inverting the in-plane oriented spins in the antiskyrmion 
rows (see circles in Fig.~\ref{fig:unitCell}b), the magnetic state obtains the rotational sense favored by the DMI. 
The resulting spin structure contains two
skyrmions in the unit cell (\fig \ref{fig:unitCell}b) and is denoted as
double-skyrmionic lattice (d-SkX).
The d-SkX state is even lower in total energy than the multi-Q state (Fig.~\ref{fig:mgkm}).
There are two new types of substructures in the d-SkX which we refer to as pseudo-skyrmions (pS), and they substitute the antiskyrmion building blocks. 

In the d-SkX the TOM of the skyrmion plaquettes is 0.25~$\mu_{\rm B}$ and thus
similar to that in the multi-Q state.
For the pseudo-skyrmion rows the TOM is smaller and inverted with respect to the corresponding antiskyrmions in the multi-Q 
state (cf.~Figs.~\ref{fig:unitCell}a,b), leading to a checkerboard antiferromagnetic TOM order. The total TOM is still compensated as the 
up-S and down-S and also the up- and down-pseudo-skyrmions compensate each other.

Next we consider a spin texture which comprises only one skyrmion substructure per
unit cell (Fig.~\ref{fig:unitCell}c), reminiscent of the nanoskyrmion lattice in the Fe monolayer on Ir(111)~\cite{Heinze2011}. This single-skyrmionic lattice (s-SkX)  
is obtained by flipping the $x$-component of four spins in the d-SkX lattice
(see circles in Fig.~\ref{fig:unitCell}c). 
The s-SkX contains one up-S and one down-pS and the resulting TOMs form a checkerboard pattern. In contrast to the previous two spin textures (multi-Q  and d-SkX) the s-SkX exhibits 
a net TOM of
about 0.24~$\mu_{\rm B}$.
If all spins of the displayed s-SkX are inverted an energetically degenerate state is obtained with
a total TOM in the opposite direction, similar to the case of the triple-Q state \cite{Nickel2023}. However, in our DFT calculations the s-SkX is energetically unfavorable with
respect to both the multi-Q state and the d-SkX (Fig.~\ref{fig:mgkm}) \footnote{Note, that it is not possible to construct a nanoskyrmion lattice which contains
only skyrmions as in Fe/Ir(111) \cite{Heinze2011}}.
Since the spin spiral minima are at larger values of 
$|\mathbf{q}|$ for Fe/Ir-3/Re(0001)
(Fig.~\ref{fig:mgkm}) the unit cell of the multi-Q state is smaller than in Fe/Ir(111) which leads to the shared spins of the skyrmion and antiskyrmion
substructures (Fig.~\ref{fig:unitCell}a). 

To understand the origin of these orbital moments emerging even without SOC, we compare the TOM calculated via DFT with an atomistic model. Within this
atomistic model, the TOM at site $i$, $\mathbf{L}_i^{\rm TO}$,
can be related to the scalar spin chirality \cite{Grytsiuk2020}
\begin{equation}
    \mathbf{L}_i^{\rm TO} = \sum_{(jk)} \kappa^{\rm TO}_{ijk} \chi_{ijk} \pmb{\tau}_{ijk}
\end{equation}
where $i, j, k$ denote neighboring lattice sites and the sum $(jk)$ is over all pairs of neighboring lattice sites of site $i$.
For the investigated lattices $\pmb{\tau}_{ijk}=\hat{ \mathbf{z}}$ is the unit vector along the direction
perpendicular to the surface.

The scalar spin chirality is defined as $\chi_{ijk} = \mathbf{s}_i \cdot (\mathbf{s}_j \times \mathbf{s}_k)$ and 
only the spin structure is needed for its calculation. This makes 
Eq.~(1) appealing to determine the TOM.
However, the topological orbital susceptibility $\kappa^{\rm TO}_{ijk}$ 
depends on the electronic structure,
 whole lattice $\kappa^{\rm TO}_{ijk}=\kappa^{\rm TO}$. For further details see Supplementary Note 3.
and to obtain it from our DFT
calculations of the $\mathbf{L}_i^{\rm TO}$,
we assume it to be independent of the chosen triangle $(ijk)$,
i.e.~$\kappa^{\rm TO}_{i}=\kappa^{\rm TO}_{ijk}$. Indeed, for all three investigated lattices, we find that $\kappa^{\rm TO}_i$ is nearly independent of the site $i$ of the Fe atom,
which is consistent with a nearly constant Fe spin moment on all lattice sites (see Supplementary Figure S2).
The values obtained for the three different skyrmionic lattices show only a very small variation among them with $\kappa^{\rm TO}_{i} = 0.019 \pm 0.002 \, \mu_{B}$ (see Supplementary Table 2 for all values).
The calculated TOMs at the individual sites obtained within the atomistic model considering a constant orbital susceptibility agree very well with the ones calculated by DFT. This demonstrates that
the spatial variation of the TOM originates mostly from the change of the spin structure, rather than from the change in the electronic structure, and that the atomistic model already gives a good approximation of the TOM.
The nanoskyrmion lattice in Fe/Ir(111) \cite{Heinze2011} shows a similar behaviour, as the TOM is vanishing for the multi-Q state and non-vanishing if the skyrmionic structures are isolated by a scissor operation (for these data see Supplementary Note 4, Supplementary Figure 3 and Supplementary Figure 4).

\begin{figure}
\includegraphics[width=0.99\linewidth]{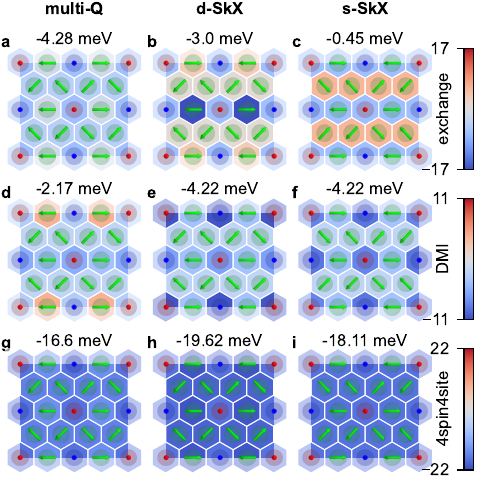}
\caption{\textbf{Energy decomposition of magnetic interactions.}
Energy contributions of
\textbf{a, b, c} the exchange interaction, 
\textbf{d, e, f} the DMI, and \textbf{g, h, i} the \foursite interaction obtained in the 
atomistic spin model for the multi-Q state (left row), the 
d-SkX lattice  (middle row), and the s-SkX lattice  (right row). All energies are displayed with respect to the FM state. The interactions of all contributing shells of neighbors have been added. A scale bar is given
for each interaction and above each plot the sum over all atoms in the unit cell is given. All energies are given in $\text{meV}$ per atom.}
\label{fig:energy}
\end{figure}

\begin{table}[]
    \centering
    \begin{tabular}{c c c c c } \hline \hline
        & \multicolumn{2}{c}{DFT} & \multicolumn{2}{c}{spin model} \\
         & w/o SOC & w/SOC & w/o SOC & w/SOC \\ \hline
    multi-Q state     & $-26.7$ & $-29.2$ & $-22.6$  & $-23.0$ \\
    d-SkX lattice  & $-27.2$ & $-31.2$ & $-24.4$ & $-26.2$ \\ 
    s-SkX lattice  & $-23.4$ & $-27.0$ & $-20.5$ & $-22.0$ \\ \hline \hline
    \end{tabular}
    \caption{\textbf{DFT energies vs.~atomistic spin model.}
    Total energies of the three atomic-scale spin lattices considered for 
    Fe/Ir-3/Re(0001) (cf.~Fig.~\ref{fig:unitCell})
    with respect to the FM state with and without (w/o) SOC. 
    The total energies are given as calculated by DFT and obtained using the
    atomistic spin model.
    All energies are given in meV/Fe atom.}
    \label{tab:latticeEnergies}
\end{table}

\textbf{Atomistic spin model.}
The DFT total energies calculated with and without SOC for all three lattices are given in Tab.~\ref{tab:latticeEnergies}. The order among the three states is the same with and without SOC, but the values show that the s-SkX and the d-SkX states gain more energy due to SOC than the multi-Q 
state.
For comparison the energies calculated by an atomistic spin model are also displayed in Tab.~\ref{tab:latticeEnergies}. The spin model has been parameterized by the DFT 
data shown in \fig \ref{fig:mgkm} 
and includes pair-wise and higher-order exchange, DMI, magnetocrystalline anisotropy energy (MAE) and anisotropic symmetric exchange (ASE)
(for details about the parameterization see Supplementary Note 1). The total energies 
of the spin model show the same trend as the DFT calculations. 

For the exchange, DMI and \foursite interaction, we present the energy contributions per Fe atom in \fig \ref{fig:energy} (for the other interactions see Supplementary Note 5 and Supplementary Figure 5). 
Among the three spin structures, the multi-Q state  exhibits the lowest exchange energy (\fig \ref{fig:energy}a) which is due to the construction as a superposition of spin spirals close
to the energy minima of the dispersion (cf.~Fig.~\ref{fig:mgkm}). Therefore, a flip of some spins increases the exchange energy  as seen for the d-SkX (\fig \ref{fig:energy}b). 
The s-SkX  is the least favoured lattice due to exchange interaction (\fig \ref{fig:energy}c). Regarding the DMI the multi-Q state  shows two rows with an opposite rotational sense with positive and negative contributions
(\fig \ref{fig:energy}d).  By flipping two spins in the unfavorable row (bottom and top row in \fig \ref{fig:energy}d) the rotational sense is switched and the DMI energy can be reduced. Therefore, the d-SkX and the
s-SkX optimize the DMI energy (\fig \ref{fig:energy}e,f) with the same rotational sense in each row compared to the multi-Q state.

All three spin lattices gain significant energy compared to the FM state
due to the \foursite interaction (Fig.~\ref{fig:energy}g-i).
Regarding the \foursite interaction the flip of two spins in the bottom row reduces the energy of the d-SkX with respect to the multi-Q state.  The in-plane rotation
of spins from the d-SkX to the s-SkX is slightly unfavorable 
in terms of the
\foursite interaction. The role played by the \foursite interaction is very similar to that in the nanoskyrmion lattice of Fe/Ir(111) \cite{Heinze2011}.
The d-SkX is energetically most favorable among all considered spin lattices. 
It gains energy due to
DMI and \foursite interaction with respect to the multi-Q state and due to
exchange and \foursite interaction with respect to the s-SkX.

\begin{figure}
\includegraphics[width=0.99\linewidth]{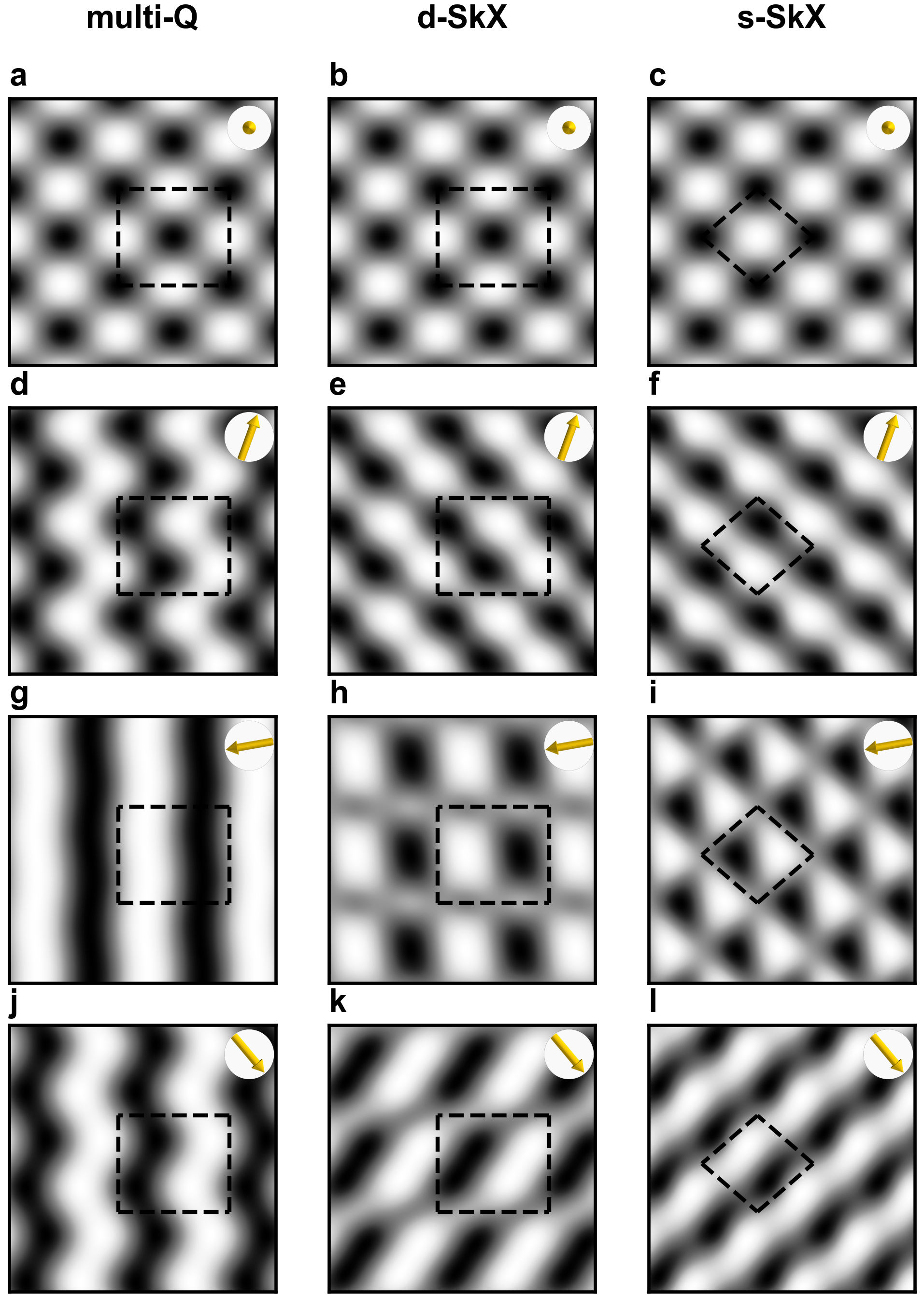}
\caption{\textbf{SP-STM simulations. } Simulated images are shown for \textbf{a, d, g, j} the multi-Q state, 
\textbf{b, e, h, k} the d-SkX, and 
\textbf{c, f, i, l} the s-SkX. For each spin lattice the simulated 
SP-STM image 
is displayed for an out-of-plane orientation of the tip magnetization (top row) and for
three in-plane orientations of the tip with an angle of $120 \, ^{\circ}$ between the different orientations (three lower rows). The 
tip magnetization direction is indicated in the top right corner of each
panel by a yellow arrow
and the dashed lines mark the magnetic unit cell as shown in \fig \ref{fig:unitCell}.}
\label{fig:stmSimu}
\end{figure}

\textbf{SP-STM simulations.}
For a direct comparison of the three spin lattices (Fig.~\ref{fig:unitCell}) with the experimental data (Fig.~\ref{fig:exp}), we simulated SP-STM images using the spin-polarized generalization of the Tersoff-Hamann model~\cite{Tersoff1985,Wortmann2001,Heinze2006}. For an out-of-plane tip magnetization as shown in \fig \ref{fig:stmSimu}a-c all three lattices have the same square contrast. These SP-STM simulations are in good agreement with the experimental contrast observed with tip\#1 on all three islands in \fig \ref{fig:exp}b. At the same time it becomes evident, that an out-of-plane tip is not suitable to distinguish between the three different proposed spin lattices. 
In contrast, different patterns are observed for the three different magnetic states when an in-plane magnetized tip is used, as evident from the SP-STM simulation in Fig.~\ref{fig:stmSimu}d-l. For each spin lattice a set of three different in-plane orientations of the tip magnetization have been used that are linked by $120^{\circ}$ angles with respect to each other, such that they correspond to a fixed tip magnetization on three different rotational domains of the spin structure as observed in the SP-STM measurements for islands A, B, and C (Fig.~\ref{fig:exp}c). 

One characteristic feature observed in the experimental SP-STM images are the stripes visible on island C. We find a similar pattern in our SP-STM simulations for the d-SkX and the s-SkX for an in-plane tip magnetization that is roughly pependicular to the stripes, see \fig \ref{fig:stmSimu}k,l. However, such stripes are never observed for the multi-Q state, regardless of the tip magnetization direction. In contrast, for the multi-Q the simulations show wavy or straight stripes along the vertical direction for all in-plane tip magnetization directions, see \fig \ref{fig:stmSimu}d,g,j. Because this kind of pattern is not in agreement with the experimental results we rule out the presense of this multi-Q state in our system.
The patterns for the d-SkX and the s-SkX resulting from the rotated tip magnetization directions in \fig \ref{fig:stmSimu}e,f,h,i can all be recognized in the experimental images of Fig.~\ref{fig:exp}c, which makes a distinction between those two states challenging. In particular the local appearance of a rectangular unit cell in island A, which is seen for both in-plane and out-of-plane magnetized tip, is likely due to a beating of magnetic and atomic pattern and not a signature of the magnetic structure as in \ref{fig:stmSimu}h.  

\subsection{Discussion}
We have found a spontaneous atomic-scale skyrmionic lattice as the magnetic ground state of an Fe
monolayer on three atomic Ir layers on the Re(0001) surface. Our first-principles calculations reveal sizeable topological orbital magnetic moments with antiferromagnetic alignment for the different proposed spin textures which can be explained by the local scalar spin chirality due to the non-collinear spin order in the skyrmion lattices. 
The topological orbital susceptibility obtained via DFT is nearly constant for all Fe atoms in the
lattice and is nearly independent of the
considered spin lattice.
This makes the atomistic model a very good approximation for the topological orbital moments. We find that Fe based film
systems 
have a significantly larger orbital susceptibility than Mn 
based film systems \cite{Nickel2023} 
(see Supplementary Table 2) which can lead to a larger 
topological orbital magnetization.

Based on the comparison with the experiment, the multi-Q state can be ruled out as the magnetic ground state of Fe/Ir-3/Re(0001)
in agreement with the DFT calculation. For the s-SkX we find a slightly better agreement with the experimental SP-STM images, whereas the d-SkX is slightly lower in total energy within DFT. Because Re becomes superconducting below 1.7 K this zero magnetic field spin lattice is a promising candidate for future studies of emerging phenomena in atomic-scale non-coplanar magnet-superconductor hybrid systems~\cite{Bedow2020,Maland2022b,Maland2023}, including the role of topological orbital moments in proximity to a superconductor.

\vspace{7mm}

\subsection{Methods}
\textbf{SP-STM experiments. }
The samples were prepared in a multi-chamber ultra-high vacuum system. The Re(0001) was cleaned by high-temperature flashes~\cite{Ouazi2014} and Ir and Fe were evaporated from rods by electron beam heating~\cite{Kubetzka2020}. Samples were then transferred \textit{in-vacuo} into an STM that is operated at $4$~K. The tip is made of Cr bulk material. The tip magnetization direction can be changed \textit{in-situ} by gentle modifications of the tip apex. The spin-polarized contribution to the tunnel current scales with the projection of tip and sample magnetization directions, and the nano-scale magnetic order can be detected directly in constant-current imaging mode~\cite{Bergmann2014}. 

\textbf{DFT calculations.} We have performed density functional theory (DFT) calculations to investigate the magnetic ground state and the magnetic interactions in Fe monolayers on three atomic layers of Ir on the Re(0001) surface. DFT
calculations have been carried out using two different methods. We have used the all-electron code {\tt FLEUR} based on the full-potential linearized augmented plane-wave (FLAPW) method \cite{FLEUR,Kurz2004,Heide2009}. In addition, we have applied the projector augmented wave (PAW) method
as implemented in the {\tt VASP} code \cite{VASP1,VASP2}. Here we provide computational details 
of the calculations using both codes.

The {\tt FLEUR} code has been used for spin spiral calculations \cite{Kurz2004} neglecting and including spin-orbit coupling (SOC), 
and to obtain the total energies of the $uudd$ 
states and the 3Q state. The SOC contribution to the energy of spin spirals was calculated using first order perturbation theory \cite{Heide2009}. The
magnetocrystalline anisotropy energy (MAE), defined as the total energy
difference between the energy for a magnetization oriented along the in-plane
and the out-of-plane direction, has been calculated self-consistently
including SOC for the FM state \cite{Li1990}. 

For all calculations using the {\tt FLEUR} code the cut-off parameter for the basis functions has been set to $k_{\rm max} = 4.1 \, \text{a.u.}^{-1}$. The radius of the muffin-tin spheres was chosen as $2.45 \, \text{a.u.}$ for Re and $2.3 \, \text{a.u.}$ for Ir and Fe. 
The exchange-correlation (xc) functional for all calculations in {\tt FLEUR} was chosen in local density approximation (LDA) \cite{vwn}. For the spin spiral calculations with and without SOC a mesh of $(44 \times 44)$ k-points was used in the full 2D-BZ. The calculations for the 3Q state and the $uudd$ states in their respective super-cells were performed on a $(22 \times 22)$ k-point mesh, which has the same density of k-points as that used in 
the spin spiral calculations. 
For the calculations of the MAE in the FM state a mesh of $(223 \times 223)$ k-points was used. For all calculations with the {\tt FLEUR} code
asymmetric films with one Fe layer, three Ir layers, and 
six layers of Re have been used. The Fe and Ir layers have been chosen in fcc stacking on the Re(0001) surface.

Structural relaxations of Fe/Ir-3/Re(0001) have been performed using the \texttt{VASP} code in the RW-AFM state. For the relaxations the GGA xc-functional PBE \cite{Perdew1996}  and a $(15 \times 15)$ grid of k-points was used.  The multi-Q state, the d-SkX, and the s-SkX have been calculated in a supercell with 16 atoms per layer and the same number of substrate layers as in the spin spiral calculations. 
Structural relaxations of Fe/Ir-3/Re(0001) have been performed using the \texttt{VASP} code in the RW-AFM state. For the relaxations the GGA xc-functional PBE \cite{Perdew1996}  and a $(15 \times 15)$ grid of k-points was used.  The multi-Q state, the d-SkX, and the s-SkX have been calculated in a supercell with 16 atoms per layer and the same number of substrate layers as in the spin spiral calculations. 
The cut-off of the wavefunctions was set to $268 \, \text{eV}$
and a grid of $(15 \times 15)$ k-points was used
for these supercell calculations.
The LDA xc-functional by Vosko, Wilk and Nusair \cite{vwn} has been applied. Topological orbital moments have been calculated using
the {\tt VASP} code. For fcc-Fe/Ir(111) they have been obtained based on the computational setup given in Ref.~\cite{Gutzeit2023}.

\textbf{Atomistic spin model. }
The atomistic spin model applied in our work contains exchange interaction, DMI, higher-order interaction (HOI), MAE and 
anisotropic symmetric exchange (ASE). The corresponding Hamiltonian is given by

\begin{small}
    
\begin{multline}
          H  = \underbrace{- \sum_{i,j} J_{ij} (\mathbf{s}_{i} \cdot \mathbf{s}_{j} ) }_{\rm exchange} 
\underbrace{- \sum_{i,j} \mathbf{D}_{ij} (\mathbf{s}_{i} \times \mathbf{s}_{j} )}_{\rm DMI}
\underbrace{-\sum\limits_{i,j} B_{ij} ( \mathbf{s}_i \cdot \mathbf{s}_j)^2}_{\rm biquadratic} \\
    \underbrace{- \sum_{ijk} Y_{ijk} [(\mathbf{s}_i \cdot \mathbf{s}_j)(\mathbf{s}_j \cdot \mathbf{s}_k) + (\mathbf{s}_j \cdot \mathbf{s}_i)(\mathbf{s}_i \cdot \mathbf{s}_k)+
    (\mathbf{s}_i \cdot \mathbf{s}_k)(\mathbf{s}_k \cdot \mathbf{s}_j)]}_{\rm 3site-4spin}  \\
    \underbrace{-\sum_{ijkl}  K_{ijkl} [(\mathbf{s}_i \cdot \mathbf{s}_j)(\mathbf{s}_k \cdot \mathbf{s}_l) + 
  (\mathbf{s}_i \cdot \mathbf{s}_l)(\mathbf{s}_j \cdot \mathbf{s}_k) 
   -   (\mathbf{s}_i \cdot \mathbf{s}_k)(\mathbf{s}_j \cdot \mathbf{s}_l)]}_{\rm 4site-4spin}\\
   \underbrace{- \sum_i K_u (\mathbf{s}_i \cdot \hat{\mathbf{z}})^2}_{\rm MAE} 
   \underbrace{-J_{\rm ASE} \sum_{ij} (\mathbf{s}_i \cdot \mathbf{d}_{ij}) (\mathbf{s}_j \cdot \mathbf{d}_{ij})}_{\rm ASE},
\end{multline}
\end{small}

where $\mathbf{s}_i$ denotes a spin at a lattice site specified by $i$, $\hat{\mathbf{z}}$ is the unit vector perpendicular to the surface and $\mathbf{d}_{ij}$ is the normalized connection vector between the lattice sites $i$ and $j$. The interactions are sorted in shells according to the distance of the lattice sites. For all atoms of each shell the same interaction constants are assumed. For the exchange and the DMI the interaction parameters for the first ten shells have been calculated. For all other interactions only the interaction parameter of the first shell has been calculated.

\textbf{SP-STM simulations. }
The simulations of SP-STM images are based on the spin-polarized generalization of the Tersoff-Hamann model~\cite{Tersoff1985,Wortmann2001}. We used a simplified SP-STM model in which
the independent orbital approximation is applied~\cite{Heinze2006}. In this model 
only the orientation of the magnetic moments with respect to the tip magnetisation 
and the tip-sample distance are needed.

\subsection{Data availability}
All relevant data are available from the corresponding authors upon reasonable request.
\subsection{Code availability}
All relevant code is available from the corresponding authors upon reasonable request.

\subsection{Acknowledgments}
It is our pleasure to thank Yuriy Mokrousov for valuable discussions.
We gratefully acknowledge financial support from the Deutsche Forschungsgemeinschaft (DFG, German Research Foundation) via projects no.~402843438, no.~418425860, and no.~462602351, and computing time provided by the North-German Supercomputing Alliance (HLRN).

\subsection{Author contributions}
K.v.B. and A.K. performed the experimental measurements and analysed the data. 
F.N. performed the DFT calculations on Fe/Ir-3/Re(0001) and all atomistic spin simulations. 
M.G. performed the DFT calculations on Fe/Ir(111). S.H. and F.N. analysed the theoretical data. K.v.B, S.H. and F.N. wrote the manuscript with contributions from all authors.

\subsection{Competing interests}
All authors declare no competing financial or non-financial interests.

\bibliography{literature}

\end{document}